\documentclass[a4paper,12pt,oneside,reqno]{amsart}
\usepackage{hyperref}
\usepackage[headinclude,DIV13]{typearea}
\areaset{15.1cm}{25.0cm}
\usepackage{txfonts,amssymb,amsmath,amsthm,bbm,dsfont}
\usepackage{multirow}
\usepackage{tabularx}
\usepackage{ccaption}

\usepackage[utf8]{inputenc}
\usepackage{graphicx, psfrag}
\usepackage{xcolor}
\usepackage{subfigure}

\newtheorem{theorem}{Theorem}[section]

\definecolor{bbm}{RGB}{51,153,0}
\definecolor{above}{RGB}{128,0,128}
\definecolor{below}{RGB}{102,0,204}
\definecolor{cascade}{RGB}{204,0,0}
\definecolor{iid}{RGB}{153,51,0}

\theoremstyle{remark}
\newtheorem*{remark}{Remark}

\def\paragraph#1{\noindent \textbf{#1}}

\numberwithin{equation}{section}

\def\<{\langle}
\def\>{\rangle}

\def\R{{\Bbb R}}  
\def\N{{\Bbb N}}

\def\E{{\Bbb E}}

 \def \X {{\Xi}}

 \def \ba {\begin{array}}
 \def \ea {\end{array}}

\def \X {{\mathbb X}}

 \newcommand{\be}{\begin{equation}}
 \newcommand{\ee}{\end{equation}}

\newcommand{\bea}{\begin{eqnarray}}
 \newcommand{\eea}{\end{eqnarray}}
\def\TH(#1){\label{#1}}\def\thv(#1){\ref{#1}}
\def\Eq(#1){\label{#1}}\def\eqv(#1){(\ref{#1})}

 \def \1{\mathbbm{1}}

 \def \bk {{\boldsymbol k}}

  \def \bX{{\boldsymbol X}}

\def \lb {\left(}
\def \rb {\right)}

\begin{document}


\title[The Effect of Mullers Ratchet on Recessive Disorders]{Refining the drift barrier hypothesis: a role of recessive gene count and an inhomogeneous Muller`s ratchet}
 
 \author[L. La Rocca]{Luis A. La Rocca}
 \address{
 	L. La Rocca\\
	Institut for Applied Mathematics \\
	University of Bonn\\
	Endenicher Allee 60\\
	53115 Bonn, Germany
	}
\email{luis.larocca@uni-bonn.de}

\author[K. Gerischer]{Konrad Gerischer}
 \address{
 	K. Gerischer\\
	Institute for Genomic Statistics and Bioinformatics\\
	University of Bonn\\
	Venusberg-Campus 1\\ 
	53127 Bonn, Germany }
\email{kgerisch@rheinahrcampus.de}

\author[A. Bovier]{Anton Bovier}
\address{
	A. Bovier\\
	Institut for Applied Mathematics \\
	University of Bonn\\
	Endenicher Allee 60\\
	53115 Bonn, Germany }
\email{bovier@uni-bonn.de}

\author[P. M. Krawitz]{Peter M. Krawitz}
 \address{
 	P. M. Krawitz\\
	Institute for Genomic Statistics and Bioinformatics\\
	University of Bonn\\
	Venusberg-Campus 1\\ 
	53127 Bonn, Germany }
\email{pkrawitz@uni-bonn.de}
 
\date{\today}

 \begin{abstract}  
The drift-barrier hypothesis states that random genetic drift constrains the refinement of a phenotype under natural selection.
The influence of effective population size and the genome-wide deleterious mutation rate were studied theoretically, and an inverse relationship between mutation rate and genome size has been observed for many species.
However, the effect of the recessive gene count, an important feature of the genomic architecture, is unknown. In a Wright-Fisher model, we studied the mutation burden for a growing number of $N$ completely recessive and lethal disease genes.
Diploid individuals are represented with a binary $2\times N$ matrix denoting wild-type and mutated alleles. Analytic results for specific cases were complemented by simulations across a broad parameter regime for gene count, mutation and recombination rates.
Simulations revealed transitions to higher mutation burden and prevalence within a few generations that were linked to the extinction of the wild-type haplotype (least-loaded class).
This metastability, that is, phases of quasi-equilibrium with intermittent transitions, persists over $100\,000$ generations.
The drift-barrier hypothesis is confirmed by a high mutation burden resulting in population collapse. Simulations showed the emergence of mutually exclusive haplotypes for a mutation rate above 0.02 lethal equivalents per generation for a genomic architecture and population size representing complex multicellular organisms such as humans.
In such systems, recombination proves pivotal, preventing population collapse and maintaining a mutation burden below 10.
This study advances our understanding of gene pool stability, and particularly the role of the number of recessive disorders.
Insights into Muller`s ratchet dynamics are provided, and the essential role of recombination in curbing mutation burden and stabilizing the gene pool is demonstrated.
 \end{abstract}

\thanks{
This work was partially supported by the Deutsche Forschungsgemeinschaft (DFG, German Research Foundation) under Germanys Excellence Strategy GZ 2047/1, Project-ID 390685813 and GZ 2151, Project-ID 390873048 and through the Priority Programme 1590 Probabilistic Structures in Evolution.}


\maketitle

 
 \section{Introduction}
The dependency of genome size and mutation rate was first noticed by Drake and further developed by Lynch into the drift-barrier hypothesis \cite{drake1991,lynch2016,sung2012}.
The basic idea of this theory is that the refinement of a phenotype is ultimately limited by the noise of genetic drift, which is a consequence of effective population size and genome-wide deleterious mutation rate.
The effective population size, $K$,  assumes a sexual diploid population that mates randomly.
Whereas the genome-wide deleterious mutation rate is the product of the base-substitution rate for deleterious mutations per nucleotide site per generation times an effective genome size that will evolve under the drift-barrier hypothesis. 
Drake`s conjecture about an approximate constant of 0.003 [deleterious] mutations per genome per generation was based on a very small number of taxa and the first generation of sequencing technology.
More than twenty years later, Sung et al. \cite{sung2012} refined this conjecture and showed an inverse relationship between the deleterious mutation rate and genome size over multiple orders of magnitude for viruses, eubacteria, and archaebacteria \cite{lynch2016}.
In order to apply the theory also to eukaryotes, the definition of an effective genome size was introduced, and the size of the coding DNA was used as a proxy.
With the latest data from studies on the human population, the base substitution rate for humans could be further specified, and the effective genome size could encompass any region where deleterious mutations may occur.
Thus, certainly the exome, but most probably also other non-coding elements such as enhancers.

\begin{figure}[ph]
	\centering
	\includegraphics[width=\textwidth]{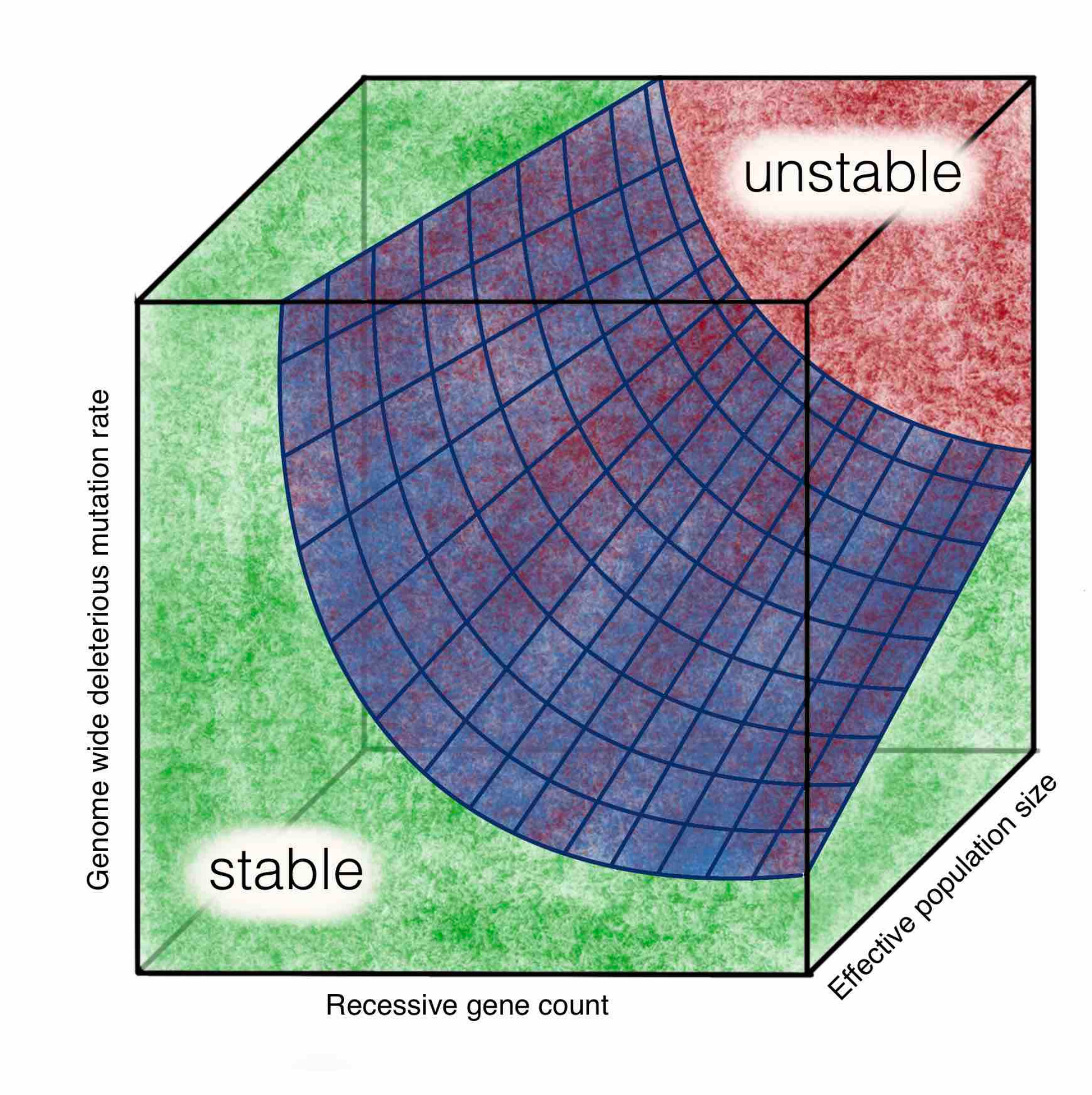}
	\caption{\textbf{A shchematic representation of the Drift-Barrier in a three-dimensional parameter space.} Below the hyperplane, populations can exist stably, whereas above it, the risk of extinction becomes too high. While the effect of the population size has already been described we observe an exponentially decreasing effect of the recessive gene count (see Figure \ref{fig:2}). The combined effect has not yet been investigated. This is merely a sketch to visualize the extension of the drift-barrier hypothesis.}
	\label{fig:0}
\end{figure}

While the drift-barrier model was originally based on empirical results, there has also been extensive theoretical work to understand the stochastic nature of this phenomenon. 
In simple terms, the dynamics of a gene pool are either stable or unstable, depending on which side of the barrier the system operates which is defined by its parameter settings (Figure \ref{fig:0}).
Most simulations in that field are based on the discrete Wright-Fisher Model or adaptive dynamics and both frameworks were shown to yield comparable results  \cite{larocca2024}.
In our work, we consider a deleterious mutation rate $\mu$, that is the number of lethal equivalents introduced per generation \emph{de novo} into the recessive genes of a gamete.
As an indicator of the fitness of a population or their gene pool, mutation burden $B(t)$ and disease incidence $P(t)$ were studied for populations of constant size $K$ over a total time span of $100\,000$ generations  \cite{henn2015}.
The proportion of diploid individuals in the population that are biallelic for such lethal equivalents in at least one gene is the incidence or prevalence rate $P(t)$.
The number of pathogenic alleles in the entire gene pool is defined as the mutation burden $B(t)$.
Incidence rate and mutation burden are also the system parameters that we use to assess its stability.
All haplotypes in the gene pool can be assigned to a class indicating the number of genes that harbour pathogenic alleles.
By that definition, $c_0$ is the least loaded class containing only the haplotype without any mutations, $c_1$ consists of all haplotypes with exactly one affected gene, and so on.
Due to maximal selection ($s=1$) and total recessivity (dominance coefficient $h=0$), a pathogenic or deleterious allele can also be referred to as a lethal equivalent that prevents propagation if both haplotypes of a gene are affected.
Muller first studied a simple stochastic process for lethal mutations in haploid genomes without recombination and observed the irreversible loss of the least loaded class from the population which he described as clicks of a ratchet \cite{haigh1978,muller1964}.
By theoretical arguments, Charlesworth and Charlesworth argued that in diploid genomes the accumulation of pathogenic or lethal recessive alleles can result in the ``crystalization'' of the population, that is, the occurrence of haplotypes that are incompatible with each other \cite{charlesworth1997}.
In our work, we study the stochastic process of Muller`s ratchet in the limit of strong selection on completely recessive diploid genomes in a parameter space with a variable number of genes that operates close to the drift barrier.
We find evidence for the crystalization phenomenon when increasing the genome-wide deleterious mutation rate or gene count beyond the drift barrier.
A particular focus of our work are the inhomogeneous or metastable dynamics that follow after the extinction of the mutation-free gametes until the emergence of haploid clusters, when the population regains stability after several thousand generations \cite{bovier2015}.
Unlike in other models, we do not reach an equilibrium every time between two successive clicks \cite{mariani2020}.
In agreement with all other models, we can show that recombination weakens the selective disadvantage of a pathogenic recessive allele by exporting mutations to other members of the population \cite{charlesworth2013,dawson1999,kimura1967,kondrashov1995,kondrashov2018}.
Therefore, recombination is also the force that helps to balance genetic load to a certain degree in genomes with an increasing number of recessive genes, which we introduced as a novel parameter of the drift barrier.

\section{Methods}
Consider a diploid population where individuals are characterized by $N$ diploid genes or gene sections that when mutated can carry and express a lethal disease.
When an individual expresses one or more of such diseases it will be excluded from the mating process and thus cannot reproduce further.
If a gene segment has already mutated once, further mutations on its same haplotype are neglected.
Therefore, we can think of the genome of an individual as a $2\times N$ matrix with values in $\{ 0, 1 \}$, where a zero represents the wild type and a one indicates the presence of at least one mutation at that location.
The fitness of an individual $x \in \{0,1\}^{2\times N}$ is optimal (fitness=1) unless it carries at least one mutation at each copy of at least one gene and hence expresses the disease.
In this case the reproductive fitness is reduced to zero. 
In a classical Wright-Fisher Model with a constant population size of $K$ individuals,
in every generation for every offspring two parents are chosen according to their fitness and the offspring then inherits a combination of the parental genetic material.
The creation of the offspring's genes is influenced by two parameters:
the probability of recombination $r$ and the mutation rate $\mu$.

The recombination rate $r\in[0,1]$ denotes the probability that a potential crossover breakpoint occurs between neighbouring genes (see Supporting Information for details).
The average number of potential breakpoints is hence Binomial distributed with parameter $N-1$ and $r$.
In the course of the copying of the DNA to form a new gamete each of the two haploid sets of chromosomes is chosen independently with equal probability.
After these recombination events, each newly born individual receives a gamete from each of their parents, creating their new diploid set of chromosomes.
Lastly, \emph{de novo} mutations are added independently with rate $\mu$ at every gene on either genome.
Therefore, the diploid mutation rate is $2\mu$, and the probability of changing the wild type to a mutated site at a specific locus is approximately $\frac{\mu}{N}$.
There are no back mutations.
Due to the usually small mutation rates and comparatively large number of possible sites, it is reasonable to assume that the total number of mutations per birth is Poisson distributed with mean $2\mu$.

\subsection{Model Description}
For notational reasons, we interpret the $2\times N$ matrix $x \in \{0,1\}^{2\times N}$ as two binary numbers, where each number represents the maternal or paternal genome.
Therefore, take an integer $i \in \{0,\dotsc,2^N-1\}$ and denote by $z_i = (z_1^i, \dotsc, z_N^i) \in \{0,1\}^N$ the $N$ digits of the dual representation of $i$ with leading zeros if necessary.
Hence the values $z^i_1, \dotsc, z^i_N \in \{0,1\}$ are chosen such that $$
i = \sum\limits_{n=1}^{N} z^i_{n} \cdot 2^{n-1}
$$
The vector $z_i$ is called haploid configuration or gamete.
With this interpretation of the diploid configuration we can easily enumerate all configurations. 
For $i,j \in \{0,\dotsc,2^N-1\}$ denote by $x_{ij} \in \{0,1\}^{2\times N}$ the genetic configuration $$
x_{ij} = \binom{z_i}{z_j} = \lb \begin{array}{rrrr}
z^i_1 & z^i_2 & \dots & z^i_{N} \\
z^j_1 & z^j_2 & \dots & z^j_{N} 
\end{array} \rb \in \{0,1\}^{2\times N}
$$
and denote by $X_{ij}(t)$ the number of individuals in generation $t$ with configuration $x_{ij}$.
Further, let $\bX(t) = \lb X_{ij}(t)\rb_{0\leq i , j\leq 2^N-1}$ be the state of the population at time $t$. 
The reproductive fitness $f$ of an individual with configuration $x_{ij}$ is defined as
$$
f\lb x_{ij} \rb \coloneqq \begin{cases}
0 & \text{, if } \exists \, n = 1,\dotsc,N \colon z^i_n = z^j_n = 1 \\
1 & \text{, else} 
\end{cases}
$$
The distribution of $\bX(t+1)$ given $\bX(t)$ is multinomial with parameters $K$ and probabilities $\lb p_{ij}(t) \rb_{0\leq i , j \leq 2^N-1}$ given by
\begin{equation}\label{eq:p_ij}
p_{ij}(t) \coloneqq \sum\limits_{h, h^\prime, k, k^\prime = 0}^{2^N-1} \frac{ X_{hh^\prime}(t)f(x_{hh^\prime})  X_{kk^\prime}(t)f(x_{kk^\prime})}{T(t)^2} m_{ij}(x_{hh^\prime},x_{kk^\prime};r,\mu)
\end{equation}
with
$$
T(t) \coloneqq \sum\limits_{i,j = 0}^{2^N-1} X_{ij}(t) f(x_{ij})
$$
the total fitness of the population.
The term $m_{ij}(x_{hh^\prime},x_{kk^\prime};r,\mu)$ denotes
the probability of two configurations $x_{hh^\prime}$ and $x_{kk^\prime}$ producing an offspring with configuration $x_{ij}$. 
Note that, due to the inheritance rules described above, we obtain the following symmetries.
First, there is no distinction between the maternal and paternal genomes, and second, the order of the partners in the composition of the genome is irrelevant.
Hence for any $i,j,h,h^\prime,k,k^\prime\in \{0,\dotsc,2^{N}-1\}$ we have 
\begin{align*}
(i) & \qquad m_{ij}\lb x_{hh^\prime},x_{kk^\prime}\rb = m_{ji}\lb x_{kk^\prime},x_{hh^\prime}\rb \\
(ii) & \qquad m_{ij}\lb x_{hh^\prime},x_{kk^\prime}\rb = m_{ij}\lb x_{h^\prime h},x_{k^\prime k}\rb
\end{align*}
Subsequently, we introduce further statistics to investigate genetic phenomena within the population.
Starting with the fraction of gametes $Z_i(t)$ in the population at time $t$ with the haploid  configuration $z_i \in \{0,1\}^N$, which is defined as
$$
Z_i(t) \coloneqq \frac{1}{2K} \sum\limits_{j=0}^{2^N-1} \lb X_{ij}(t) + X_{ji}(t) \rb
$$
and $\mathbf{Z}(t) = \lb Z_{i}(t)\rb_{0\leq i \leq 2^N-1}$ the gamete distribution at time $t$. 
For a gamete $z_i \in \{0,1\}^N$ we define the \textbf{mutation burden} $b(z_i)$ as the number of lethal equivalents on that specific haploid configuration, hence
$$
b(z_i) \coloneqq \sum_{n=0}^N z_n^i
$$
By some abuse of notation we set the mutation burden of a diploid configuration $x_{ij} \in \{0,1\}^{2\times N}$ to be $b(x_{ij}) \coloneqq b(z_i)+b(z_j)$. The mean haploid mutation burden $\beta(t)$ of the population $\bX(t)$ in generation $t$ is defined as the weighted mean of $\mathbf{Z}(t)$ with weights according to $b$, hence
$$
\beta(t) \coloneqq \sum\limits_{i=0}^{2^N-1} b(z_{i})Z_{i}(t)
$$
To measure fluctuations within the haploid mutation burden of a population, we also look at the weighted variance $\sigma^2_b(t)$ of $\mathbf{Z}(t)$ with weights $b$ at time $t \geq 0$ and set 
$$
\sigma^2_b(t) \coloneqq \sum\limits_{i=0}^{2^N-1} Z_i(t) \lb b(z_{i}) - \beta(t) \rb^2
$$
When we speak about mutation burden usually, we mean the average number of mutations per individual in the population at time $t\geq 0$.
Hence the \textbf{mutation burden} $B(t)$ of the population at time $t\geq 0$ is defined as $B(t)\coloneqq 2\beta(t)$.
Moreover, we define the \textbf{prevalence} $P(t)$ as the inverse of the relative fitness $$
P(t) \coloneqq 1 - \frac{T(t)}{K} = \frac{1}{K} \sum\limits_{i,j = 0}^{2^N-1} X_{ij}(t) (1-f(x_{ij}))
$$
At every haploid locus there are two possible alleles.
The wild type $(0)$ and the mutant allele $(1)$.
For $n=1,\dotsc,N$, we define the haploid \textbf{allele frequency} $\varphi_n(t)$ of the mutant allele at locus $n$ at time $t$ as
$$
\varphi_n(t) \coloneqq \sum\limits_{i=0}^{2^N-1} z_n^i \cdot Z_i(t) 
$$ 
and the average allele frequency across all loci as $\varphi(t) \coloneqq \frac{1}{N} \sum_{n=1}^N \varphi_n(t)$.
For any $k \in \{0,\dots,N\}$, define the \textbf{haploid load class} $c_k(t)$ as the fraction of gametes at time $t\geq0$ with exactly $k$ mutated genes
$$
c_k(t) \coloneqq \sum_{i=0}^{2^N-1} Z_i(t)\mathds{1}_{\left\{b(z_i) = k\right\}}
$$
The discrete probability distribution $H(t)$ on $\{0,1,\dots,N\}$ with weights $\lb c_k(t) \rb_{k=0,\dotsc,N}$ is called the haploid load class distribution.
Note that by assigning each gamete uniquely to one of the $N+1$ disjoint load classes, we obtain $\sum_{k=0}^N c_k(t) = 1$.
The mean and variance of the haploid load class distribution are given by $\beta(t)$ and $\sigma^2_b(t)$ respectively. 
Lastly, we are interested in the level of linkage (dis-)equilibrium between different genes within the population, and we introduce the joint allele frequency $\varphi_{n,m}(t)$ of the mutated allele at the two distinct loci $n,m \in \{1,\dots,N\}$ as
$$
\varphi_{n,m}(t) \coloneqq \sum_{i=0}^{2^N-1} z_n^iz_m^i \cdot Z_i(t)
$$ 
The square coefficient of correlation between the pair of loci is defined as
$$
\rho^2_{n,m}(t) \coloneqq \frac{\lb \varphi_{n,m}(t)-\varphi_n(t)\varphi_m(t) \rb^2}{\varphi_n(t)\lb 1-\varphi_n(t) \rb \varphi_m(t) \lb 1-\varphi_m(t) \rb}
$$ and consequently $ \lb \rho_{n,m}(t) \rb_{1 \leq n,m \leq N}$ is called the \textbf{correlation matrix} between the loci at time $t\geq0$.
All notations are summarized in table \ref{tab:not} at the end of this paper.

\section{Results}
\subsection{Mutation burden beyond the drift barrier}
\begin{figure}[ph]
	\centering
	\includegraphics[width=\textwidth]{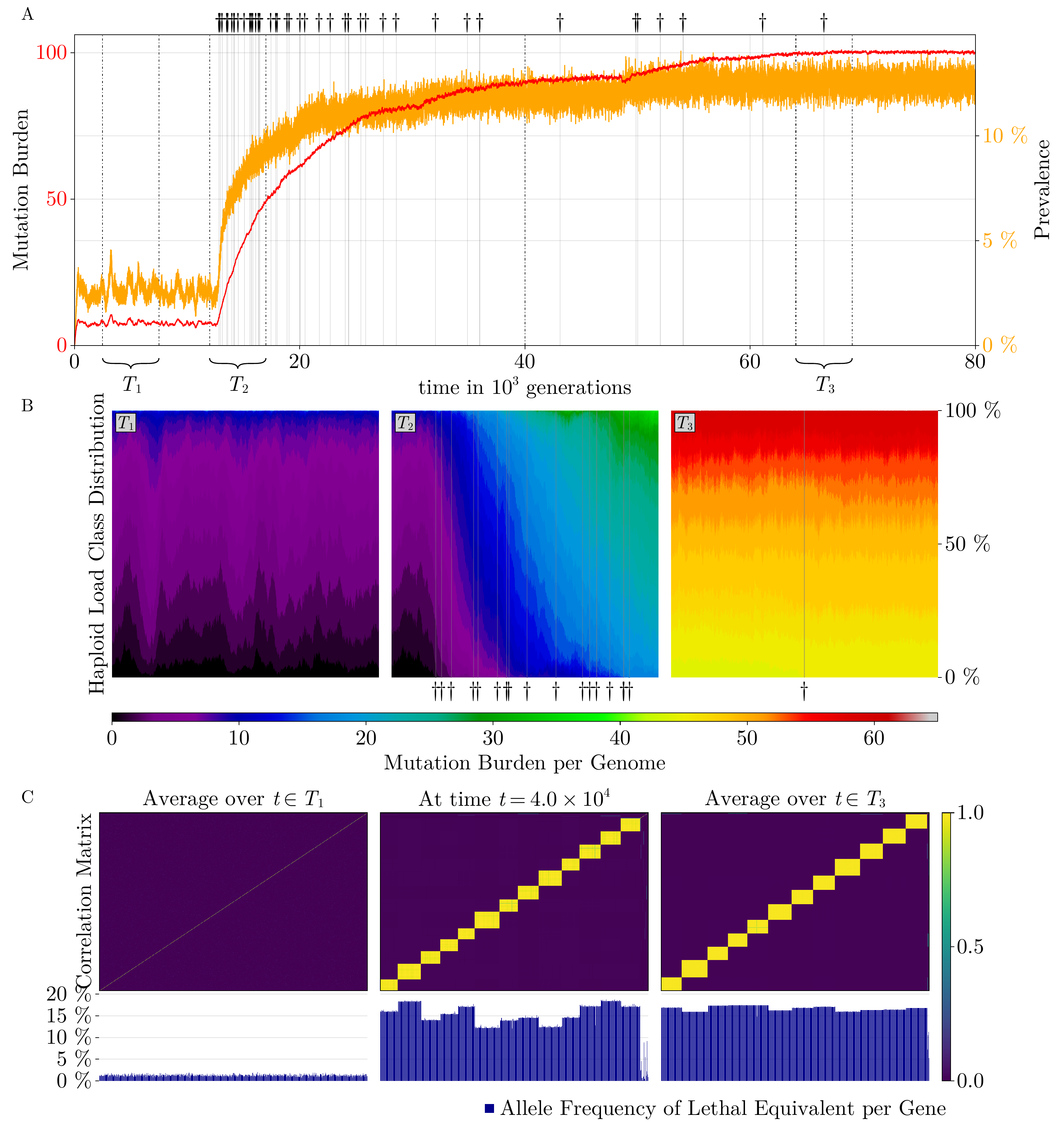}
	\caption{\textbf{Metastability of mutation burden.} For $K=10\,000$, 
$N=600$, $2\mu = 0.05$, the population operates close to the drift barrier. A) Over a time span of $100\,000$ generations, several transitions to higher levels of mutation burden and prevalence can be observed that occur within a few generations and remain constant over many generations (metastability). B) The molecular cause for a transition to a higher level is the extinction of the least loaded class $c_n$, indicated by $\dagger$ ( the first dagger is the loss of $c_0$, the class of haplotypes without any pathogenic mutation, and so on). C) The correlation or similarity matrix of the haplotypes changes over time. In the beginning, mutations are randomly distributed over the genes, and haplotypes are not correlated ($T_1$). At generation 40k, most haplotypes can already be assigned to one out of thirteen clusters with a heterozygote advantage. The cluster sizes, as well as the proportion of haplotypes assigned to clusters, increase over time, and the number of clusters decreases. In this simulation, presumably, a fixed state with twelve clusters was reached at $T_3$.}
	\label{fig:1}
\end{figure}

The initial model that we analyzed, consisted of a genome with $2\mu=0.05$, $N=600$ recessive genes, and an effective population size of $K=10\,000$ diploid individuals.
Exact values are not known, but empirical data suggest that the orders of magnitude should be correct \cite{martin2018,xiao2021,narasimhan2016,gravel2011,henn2015}.
In the simulation shown in Figure \ref{fig:1}, mutation burden and prevalence remain constant over many generations.
In this equilibrium, the count of deleterious mutations that are added to the gene pool, is equal to the number of such variants that are removed due to the affected individuals who do not procreate.
To calculate these equilibria from the differential equations coming from the multinomial sampling in \eqref{eq:p_ij} is very hard.
Already giving explicit formula for the probabilities $\lb p_{ij}(t) \rb_{i,j=0,\dotsc,2^N-1}$ is quite challenging and in many cases does not give much insight.
However in the simplest case for $N=1$ explicit calculations are feasible (see Appendix).
These match the considerations of Nei \cite{nei1968} for complete recessive lethals, which result in an allele frequency that is equal to the square root of the mutation probability.
Extending these results to $N>1$, assuming that in equilibrium the allele frequencies across all loci are equal yields an equilibrium allele frequency of
$$
\varphi = \sqrt{1-e^{-\frac{\mu}{N}}},
$$
per gene, since the mutation probability for every individual gene with $N$ total genes is equal to $1-e^{-\frac{\mu}{N}}$.
This leads to a mutation burden for an individual with $2N$ genes of
$$
B = 2N\varphi = 2N\sqrt{1-e^{-\frac{\mu}{N}}}
$$
and a prevalence of
$$
P = 1-(1-\varphi^2)^N = 1-e^{-\mu}.
$$
Note that the prevalence is independent from the number of loci $N$ and equals the probability that a mutation appears on a gamete at birth.
In Figure \ref{fig:3} A we see that in the case of no recombination ($r=0$) the simulations match these considerations.
In the case of full recombination ($r=1$) however the allele frequency and hence also the haploid mutation burden is lower that what was expected by Nei \cite{nei1968}.
However in the absence of recombination, these equilibria, while they may remain stable for several thousand generations, are fragile. 
For example, in Figure \ref{fig:1} at around generation 20k we observe a transition to a higher mutation burden and prevalence within a few generations.
Further transitions follow over a larger time frame in a stochastic manner.
On the level of haplotypes, these transitions are a consequence of the extinction of the least loaded classes, which has also been referred to as clicks of  Muller`s ratchet (Figure \ref{fig:1} B).
After the extinction of the $c_0$ class the new least loaded class $c_1$ of gametes with exactly one lethal equivalent is left without influx, but rather looses gametes due to \emph{de novo} mutations.
This would lead to a rapid fixation of one mutation within the population, even under weaker selection coefficients \cite{charlesworth1997}.
Fixation however is not possible without the extinction of the whole population for recessive lethals.
Instead we observe that the extinction of the mutation free gamete sets off a cascade of extinction events that is only reassured by the formation of clusters of similar haplotypes that are mutually exclusive. 
This phenomenon of ``crystalization'' was already predicted for low dominance coefficients by Charlesworth and Charlesworth \cite{charlesworth1997}.
In the period after the population has stabilized, there can be instances of further genes being incorporated into one of the clusters.
This may or may not be accompanied by the extinction of the least loaded class.
Moreover, there can be a reduction in the number of clusters as they compete with each other.
This results not only in an increase in mutation burden but also in a rapid rise in prevalence (see Figure \ref{fig:1} at around 50k).

\subsection{Influence of recessive gene count on metastability}
\begin{figure}[ph]
	\centering
	\includegraphics[width=\textwidth]{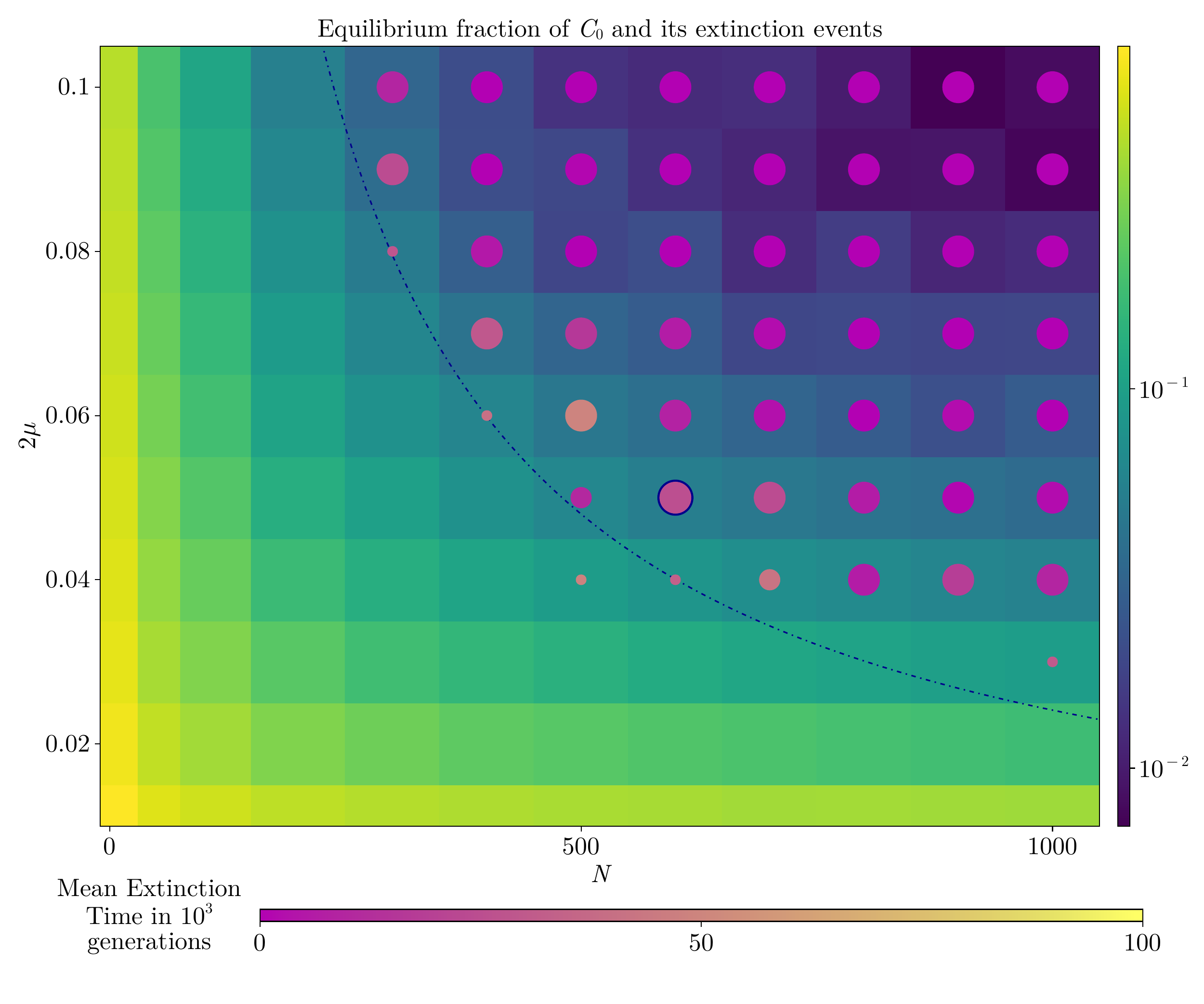}
	\caption{\textbf{Influence of recessive gene count on the drift barrier.} If the extinction of the least loaded class occurs within $100\,000$ generations, metastability occurs. The grid visualizes the outcome of three iterations for each parameter combination, and the radius of the circles in the squares indicates the probability of extinction. E.g. for a deleterious mutation rate of $2\mu = 0.05$ and $N = 600$ genes, the least loaded class $c_0$ died out in all three simulations. In contrast, for $N=200$, this event was not observed despite the same mutation rate. This indicates that the phase transition depends not only on the genome-wide mutation rate $\mu$ but also on $N$, and the recessive gene count becomes an additional parameter of the genomic architecture. The dotted line indicates the 99\% quantile of extinction events.}
	\label{fig:2}
\end{figure}

The metastability that we observed for mutation burden and prevalence occurred on the time scale of 100k generations in all of our simulations but at different time points, indicating the stochastic nature of this process (Muller`s ratchet).
Under natural conditions, the transitions to a much higher mutation burden in the gene pool result in the extinction of the population for the combination of $\mu$ and $K$, which would be in accordance with the drift barrier hypothesis (see Supporting Information for details).
In the following experiments, we aimed to characterize the interplay of $\mu$ and the number of recessive genes.
We, therefore, counted how often and when the first click of Muller`s ratchet, which is the loss of the $c_0$ class, happened (Figure \ref{fig:2}).
We found that for a given $K$ the drift barrier depends not only on $\mu$ but also on the recessive gene count $N$.
As long as the proportion of $c_0$ haplotypes in the gene pool remains above roughly $7.3\cdot10^{-3}$ a transition to higher mutation burden and prevalences is unlikely to occur within 100k generations (dotted line in Figure \ref{fig:2}).
For a genome with $N=400$ recessive genes, there were still enough haplotypes in the $c_0$ class at a mutation rate of $2\mu = 0.03$.
However, when the gene count increased to $N=1000$, mutations started to accumulate and $c_0$ died out unless the mutation rate was lowered to roughly $2\mu < 0.015$.
This suggests that Drake`s Rule can also be formulated as the minimal proportion of the haplotypes without deleterious mutations, $c_0$, in the gene pool that is required to avoid Muller`s ratchet.

\subsection{Recombination can avoid the extinction of the least-loaded class}
\begin{figure}[ph]
	\centering
	\includegraphics[width=\textwidth]{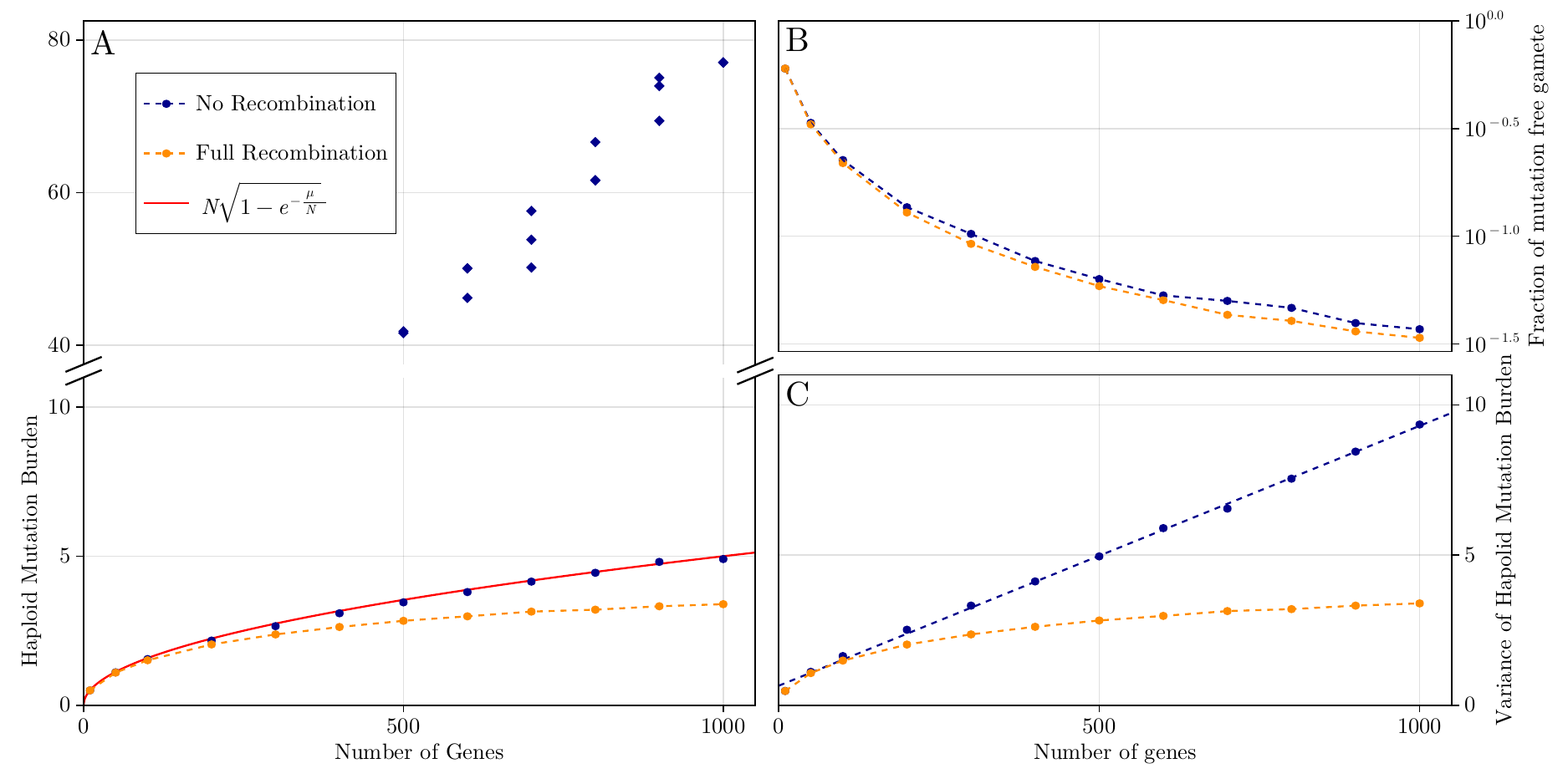}
	\caption{\textbf{Recombination can effectively control mutation burden for higher gene counts.} A) The number of recessive genes $N$ and the genome-wide deleterious mutation rate $\mu$ affect the drift barrier. For $2\mu = 0.05$ and $N = 500$, metastability would already occur (see Figure \ref{fig:2}), and the average mutation burden after $100\,000$ generations deviates from the beginning of the simulations. The average haploid mutation burden before and after the transition is indicated by circles and diamonds. The mutation burden before any transition and without recombination can be described by $N \sqrt{1-e^{-\frac \mu N}}$, and the variance increases linearly (C). However, beyond the drift barrier, the system becomes unstable, and mutation burdens after the transitions can differ. With recombination, the mutation burden in more than 500 recessive genes can remain below 10, which is considered an important threshold for lethal equivalents in the mutation-selection equilibrium. Likewise, the variance of the haploid mutation burden increases considerably slower with recombination (B).}
	\label{fig:3}
\end{figure}

Beyond the drift barrier, the gene pool quickly acquires deleterious mutations that can eventually result in a collapse of the entire population.
In finite populations this stochastic mechanism can be counteracted by amphimixis, that is sexual reproduction involving the fusion of two different gametes, that have to undergo meiosis \cite{keightley2006,kondrashov2018,muller1964,otto2001}.
During meiosis, recombination can occur that counteracts linkage disequilibrium (LD) between deleterious mutations in different genes by negative selection.
In our simulations, the recombination rate is the probability of a crossing over between genes.
That is, for a recombination rate of $r=0$, either the grandmaternal or grandpaternal haplotype is transmitted.
In contrast, for a recombination rate of $r=1$, the resulting haploid genome of the gamete would be a random sequence of the ancestral genes.
We observed that in a population without recombination, metastability occurs for $2\mu = 0.05$ around $N = 500$ genes.
That means the mutation burden in the gene pool after $100\,000$ generations deviates strongly from the beginning.
The introduction of recombination, however, is able to control the mutation burden effectively and keep it below 10 for $N = 1000$, and probably above (Figure \ref{fig:3}).
Even in the initial equilibrium stage during which the mutation free gamete is present, the frequency of the mutated allele is reduced by recombination.
This inevitably is followed by an - albeit small - reduction in the size of the least loaded class (Figure \ref{fig:3} B).
And yet recombination prevents the extinction of the mutation free gamete, because of two reasons.
First, recombination lowers the fluctuation within the population.
On the one hand for full recombination the haploid load class distribution $H$ is a \emph{Poisson} distribution, the variance of the haploid mutation burden stays comparable low for large $N$, like the expectation.
Whereas on the other hand we observe a linear growth in $N$ for the variance in the case of no recombination (Figure \ref{fig:3} C).
These higher variances due to the absence of recombination, similar to a reduction of the population size, favour the extinction of the least loaded class due to natural fluctuations.
Second - and more important - recombination can restore the mutation free gamete after it got lost.
To see that denote by $g_r:\{0,1\}^{2\times N} \rightarrow [0,1]$ the probability that a given genetic configuration with recombination rate $r\in[0,1]$ produces a mutation free gamete.
Here, we consider only the effect of recombination during gamete formation, without the influence of \emph{de novo} mutation, as they are added only to the newly formed diploid individual.
Then
$$
g_1(x_{ij}) = \begin{cases}
0 & \text{ , if } f(x_{ij}) = 0 \\
\left(\frac{1}{2}\right)^{b(x_{ij})} & \text{ , else.}
\end{cases}
$$
In particular note, that even if at some point in time $c_0 = 0$ the class of mutation free gametes may still have an influx with positive probability.
That is true whenever $r>0$.
Only in the absence of recombination and only segregation we get
$$
g_0(x_{ij}) = \begin{cases}
1 & \text{ , if } i+j = 0 \\ 
\frac{1}{2} & \text{ , if } i\cdot j = 0 \text{ and } i+j>0 \\
0 & \text{ , else.}
\end{cases}
$$
In that case, if at some point $t_\dagger > 0$ the class of mutation free gametes goes extinct by natural fluctuations it will stay extinct for all $t \geq t_\dagger$ and every new gamete will carry at least one mutation, which is known as a click of Mullers ratchet \cite{muller1932}. Theoretical models widely acknowledge that even low recombination rates can decelerate the accumulation of mutations \cite{felsenstein1974,muller1964}.

\section{Discussion}
In this work we showed that the recessive gene count is another parameter that is required to decide whether a population can reach a mutation selection balance or in other words whether it is able to operate on the stable side of the drift barrier (Figure \ref{fig:0}).
In fact, we observed the complex dynamics of the mutation burden beyond the barrier for the first time by coincidence in our previous work, in which we studied the effect of different mating schemes and demographic histories on the recessive disease risk and incidence rates \cite{larocca2024}.
We expected that selection would be sufficient as an opposing force to counteract the effect of deleterious mutations and genetic drift in the full range of population sizes that we simulated.
However, for certain parameter settings, particularly many recessive genes that are linked on the same chromosome, we noticed a metastability of the mutation burden and investigated this phenomenon further.
Prior to us, the puzzling observation of a population collapse had been made by theoretical biologists studying molecular evolution and e.g in the book the ``Crumbling Genome'', Kondrashov theorised about the origins of sexual mating \cite{kondrashov2017}.
Amphimixis, that is reproduction of a diploid multicellular organism by means of haploid sperm or egg cells, is just one possibility how Nature implemented genetic recombination \cite{kondrashov2018}.
By recombination, a reconstruction of the wildtype sequence becomes possible, even if lethal equivalents affect multiple recessive genes.
Kondrashov estimated that the average rate of ``contamination'' cannot surpass 10 in a human genome, or otherwise mutation-selection balance would not be sustainable \cite{kondrashov2017}.
Kondrashov defined contamination as the sum of all heterozygous pathogenic variants in recessive genes weighted by their selection coefficient and coined the term ``muller'' for this unit.
10 mullers would be 10 lethal equivalents per genome, or 100 pathogenic variants with a selection coefficient of $s=-0.1$, and so on.
In retrospect, we can confirm that passing this threshold in our previous work caused the metastability, which also explains the reproducibility issues that we had for different seeds in our simulations.
However, with recombination we were able to keep the mutation burden below 10 effectively for $N=1\,000$, $2\mu =0.05$, and a population size of $K = 10\,000$.

In the present work, we aimed at approaching the drift barrier by numerical simulations for wide parameter ranges in all three dimensions. We did so by means of adaptive dynamics that behave equivalently to classical Wright-Fisher population genetics. 
We could also confirm the emergence of multiple mutually exclusive haplotypes, and for smaller numbers of genes.
This ``crystallization'' into two complementary segregating haplotypes, is a phenomenon predicted by Brian Charlesworth after studying earlier works from P\`{a}lsson, et al. \cite{charlesworth2012,palsson1999,palsson2001}.
In fact, we see our work as a new part of a long sequence of findings ever since John Haigh established a mathematical model in 1978 to quantify certain effects of Muller's ratchet \cite{haigh1978}.
Particularly, the question of the click rate - that is, the speed at which successive least loaded classes become extinct - is of great interest. Recently, good approximations depending on effective population size have been achieved using diffusion approximation \cite{etheridge2007}.
However, all these models observe time-homogeneous click rates.
Interestingly, we observe highly inhomogeneous click rates and we hypothesized that this is due to the higher complexity of the genomic architecture that we modelled.
This makes techniques like diffusion approximation or coalescent approaches from previous studies inapplicable \cite{gonzalezcasanova2023,audiffren2013}.
Indeed, after simplifying our model to only consider the number of deleterious mutations per gamete and assuming these mutations are always uniformly distributed across the genome in each step, we observed a homogeneous click rate again.
Thus, the position of mutations is crucial, making mathematical analysis very challenging.
The inhomogeneity is an interesting finding from a mere stochastic point of view and could be motivation enough to analyse the timespan between clicks and determine their distribution.
Furthermore, for evolutionary biology it might be interesting to focus not only on the clicks of Muller's ratchet but also on the addition of recessive genes to a cluster - which correlates with an increase in mutation burden - and the extinction of clusters - which correlates with an increase in prevalence.
By this means our model can also contribute to the understanding of multilocus dynamics, that have recently been shown to the associative overdominance and background selection \cite{gilbert2020}.
Providing these measures in terms of the order of effective population size would be a significant challenge.
We share Charlesworth's hypothesis that this evolution stops only when either the entire population becomes extinct or two exclusive clusters emerge, each encompassing the entire set of genes.

\section{Code Availability}
The code that supports the findings of this study has been deposited in an open-access repository  and can be accessed via the following link: \url{https://doi.org/10.5281/zenodo.10985649}. The repository contains detailed instructions for code usage, dependencies, and any other relevant information to facilitate reproducibility of the results.

\section{Further Engagement and Presentation}
For those interested in a visually guided introduction to our main findings, I recently gave a short 15-minute talk at the European Society of Human Genetics (ESHG) conference in Berlin. This talk serves as an accessible entry point to our research, with the detailed context and insights provided in this paper. You can view the talk on YouTube via the following link: \url{https://youtu.be/QXoHY5ODkVk}

\newpage

\section{Notation}
\begin{table}[tbhp]
	\centering
	\caption{Notation used within this paper}
	\label{tab:not}
	\renewcommand{\arraystretch}{1.5}
	\begin{tabularx}{\textwidth}{|m{9,655cm}|c|}
		\hline
		Description & Notation \\
		\hline \hline
		number of diploid genes & $N \in \N$ \\
		\hline
		total / effective population size & $K \in \N$ \\
		\hline
		recombination rate & $r \in [0,1]$ \\
		\hline
		haploid genome mutation rate & $\mu \in \R_{\geq 0}$ \\
		\hline
		individual / configuration / genome & $x_{ij} \in\{0,1\}^{2\times N}$ \\
		\hline
		gamete / haploid configuration / haplotype & $z_{i} \in\{0,1\}^{N}$ \\
		\hline
		integers which binary representation represents the haploid genome & $i,j = 0,\dotsc,2^N-1$ \\
		\hline
		digits of the binary numbers representing the state of one haploid gen & $z_n^i \in \{0,1\}$ \\
		\hline
		state of the population at time $t \in \N$ & $\bX(t) = \lb X_{ij} \rb_{i,j=0,\dotsc,2^N-1} \in \N^{2^N}$ \\
		\hline
		number of individuals with configuration $x_{ij}\in\{0,1\}^{2\times N}$ at time $t\in\N$  & $X_{ij}(t) \in \{0,\dotsc,K\}$ \\
		\hline
		proportion of gametes with haploid configuration $z_{i}\in\{0,1\}^{N}$ at time $t\in\N$  & $Z_{i}(t) \in [0,1] $ \\
		\hline
		probabilities of multinomial distribution of $\bX(t+1)$ given $\bX(t)$  & $\lb p_{ij} \rb_{i,j=0,\dotsc,2^N-1} \in [0,1]^{2^N}$\\
		\hline
		reproduction probabilities - probability that a paring of $x_{hh^\prime}$ and $x_{kk^\prime}$ results in $x_{ij}$ under the mutation rate $\mu$ and recombination rate $r$ &$m_{ij}\lb x_{hh^\prime},x_{kk^\prime}; r, \mu \rb \in [0,1]$\\ 
		\hline
		haploid mutation burden - number of mutations of the haploid configuration $z_{i}$ & $ b\lb z_{i}\rb \in \{0,\dotsc,N\} $\\
		\hline
		relative mutation burden of the population at time $t \in \N$ & $ B(t) \in \R_{\geq 0}$ \\
		\hline
		prevalence of the population at time $t \in \N$ & $ P(t) \in [0,1]$ \\
		\hline
		local allele frequency of the mutated allele at the $n^\text{th}$ gene at time $t\in\N$ & $\varphi_n(t) \in [0,1] $ \\ 
		\hline
	\end{tabularx}
\end{table}

\newpage

\section{Figures}

\newpage

\newpage

\newpage

\newpage

\section{Appendix}
\subsection{Only one gene}
For $N = 1$ there are three distinct configurations $$
x_{00} = \binom{0}{0} \quad,\quad x_{01} = \binom{0}{1} \quad , \quad x_{11}=\binom{1}{1}.
$$ Symmetric configurations like $x_{10}$ and $x_{01}$ are indistinguishable in the sense that they have the same reproductive rates as mentioned above.
The terms $\lb m_{ij}\rb_{i,j \in \{0,1\}}$ that describe the probability that a given mating produces an offspring of type $ij$ are given by the following table 
\renewcommand{\arraystretch}{1.5}
\[ \begin{array}{ | c || c | c | c | }
\hline 
& x_{00} \times x_{00}
& x_{00} \times x_{01}
& x_{01} \times x_{01} \\ \hline \hline
m_{00} 
& e^{-2\mu} 
& \frac{1}{2}  e^{-2\mu} 
& \frac{1}{4} e^{-2\mu}
\\ \hline
m_{01} 
& 2e^{-\mu} \lb 1-e^{-\mu}\rb 
& e^{-\mu}\lb 1- e^{-\mu} \rb + \frac{1}{2} e^{-\mu} 
& \frac{1}{2} e^{-\mu} \lb 1- e^{-\mu} + \frac{1}{2} e^{-\mu} \rb   
\\ \hline
m_{11} 
& \lb 1-e^{-\mu} \rb^2 
& \frac{1}{2} \lb 1-e^{-\mu} \rb^2 + \frac{1}{2} \lb 1 - e^{-\mu} \rb  
& \frac{1}{4} \lb 1 -e^{-\mu} \rb^2 + \frac{1}{2} \lb 1 - e^{-\mu} \rb + \frac{1}{4} 
\\ \hline
\end{array} \] 
\renewcommand{\arraystretch}{1}
Since individuals with the phenotype $x_{11}$ have a reproductive fitness of zero and are thus excluded from the mating process, their probability of generating any configuration is negligible. 
A particular distribution $\bk = (k_{00},k_{01},k_{11})$ is stationary if $$
\E\left[\bX(t+1) | \bX(t) = K\bk \right] = K\bk
$$ Since $\bX(t+1) | \bX(t) \sim \mathrm{Multinomial}\left(K;p_{00},p_{01},p_{11}\right)$ we have that $$
\E\left[X_{ij}(t+1) | \bX(t) \right] = Kp_{ij} \qquad \text{,for } 0\leq i \leq j \leq 1.
$$ which yields \begin{align} \label{eq:stationary00}
k_{00} & = \frac{k_{00}^2 + k_{00}k_{01} +  \frac{1}{4} k_{01}^2}{(k_{00}+k_{01})^2} e^{-2\mu} \\
\label{eq:stationary01}
k_{01} & = \frac{k_{00}^2 + k_{00}k_{01} +  \frac{1}{4} k_{01}^2}{(k_{00}+k_{01})^2} 2e^{-\mu}\lb1-e^{-\mu}\rb 
+ \frac{k_{00}k_{01}+\frac{1}{2}k_{01}^2}{(k_{00}+k_{01})^2}e^{-\mu} \\
\label{eq:stationary11}
k_{11} & =  \frac{k_{00}^2 + k_{00}k_{01} +  \frac{1}{4} k_{01}^2}{(k_{00}+k_{01})^2} \lb 1-e^{-\mu} \rb^2 
+  \frac{k_{00}k_{01}+\frac{1}{2}k_{01}^2}{(k_{00}+k_{01})^2} \lb 1-e^{-\mu} \rb 
+ \frac{\frac{1}{4}k_{01}^2}{(k_{00}+k_{01})^2}
\end{align}
\begin{theorem}
	The unique positive solution of the system (\ref{eq:stationary00}-~\ref{eq:stationary11}) is given by
	$$
	k_{00} = \lb 1-\sqrt{1-e^{-\mu}}\rb^2, \quad k_{01} = 2\lb1-\sqrt{1-e^{-\mu}}\rb \sqrt{1-e^{-\mu}}, \quad k_{11} = 1-e^{-\mu}
	$$
\end{theorem}
\begin{proof}
	Set 
	$$
	\rho = \frac{k_{00}+\frac{1}{2}k_{01}}{k_{00} + k_{01}}
	$$
	then the system (\ref{eq:stationary00}-~\ref{eq:stationary11}) changes to
	\begin{align*}
	k_{00} &= \rho^2e^{-2\mu} \\
	k_{01} &= 2\rho e^{-\mu}\lb 1-\rho e^{-\mu} \rb \\
	k_{11} &= \lb 1-\rho e^{-\mu} \rb^2
	\end{align*}
	and we can solve for $\rho$ in
	$$
	\rho = \frac{\rho^2e^{-2\mu}+\rho e^{-\mu}\lb 1-\rho e^{-\mu} \rb}{\rho ^2e^{-2\mu} + 2\rho e^{-\mu}\lb 1-\rho e^{-\mu} \rb} = \frac{1}{2-\rho e^{-\mu}}
	$$ 
	which results in
	$$
	\rho = \frac{1\pm\sqrt{1-e^{-\mu}}}{e^{-\mu}}.
	$$ Therefore the only solutions with only positive entries is given by
	$$
	k_{00} = \lb 1-\sqrt{1-e^{-\mu}}\rb^2, \quad k_{01} = 2\lb1-\sqrt{1-e^{-\mu}}\rb \sqrt{1-e^{-\mu}}, \quad k_{11} = 1-e^{-\mu}
	$$
\end{proof}
\begin{remark}
	One can immediately see that the population in equilibrium satisfies the Hardy-Weinberg principle with frequencies for the mutated allele of
	$$
	\varphi = \sqrt{1-e^{-\mu}}
	$$
	Moreover we find the equilibrium mutation burden and prevalence for $N=1$ as $$
	\hat{B} = 2\sqrt{1-e^{-\mu}} \qquad \text{and} \qquad \hat{P} = 1-e^{-\mu}
	$$
\end{remark}

\subsection{A diploid individual based model of adaptive dynamics}\label{sec:nat_ext}
In addition to the Wright-Fisher model, we also implemented an adaptive dynamics model.
Unlike Wright-Fisher models, the latter operates with a fluctuating population size and overlapping generations.
Here, individuals reproduce and die at independent, exponentially distributed times, determined both by the individual's fitness and, in the case of the death rate, by the competitive pressure of the population.
Accordingly, it can better capture effects that lead to growth or shrinkage of the population than population models with a fixed, constant population size.
In the limit of large populations, we observe no differences however between the Wright-Fisher and the Adaptive Dynamics model for populations in equilibrium \cite{larocca2024}.

\subsubsection{Model Description}
We use a variation of the model of adaptive dynamics of Mendelian diploids studied by  P. Collet, S. M\'{e}l\'{e}ard, J. Metz et al. \cite{collet2013}. The major adaptation we make is a finite, but high dimensional genotype space $\mathcal{X} \subset \R^N$ and a more general approach on the recombination and propagation mechanism of genotypes during a mating of individuals. Here the dimension $N$ corresponds to the number of gene segments under consideration. Hence a diploid individual is characterized by its genotype $\mathbf{x} = (x_1,x_2) \in \X^2 $.
In the following we introduce the demographic parameters that encode all of biology.
We assume that these parameters are influenced by the allelic traits through the phenotypic trait.
As this dependency is symmetrical, all coefficient functions defined are also assumed to be symmetric in the allelic traits. 
\begin{itemize}
	\item[(i)] $b(x_1,x_2) \in \R_+$ : on the one hand this is the birth rate of an individual with genotype $(x_1,x_2)$ and on the other hand an individual with genotype $(x_1,x_2)$ has probabilities proportional to $b(x_1,x_2)$ to be chosen as a mate during the birth event of another individual.
	\item[(ii)] $d(x_1,x_2) \in \R_+$ :  the intrinsic death rate of an individual with genotype $(x_1,x_2)$.
	\item[(iii)] $c(x_1,x_2,y_1,y_2) \in \R_+$ :  the competition pressure from an individual with genotype $(y_1,y_2)$ exerted onto an individual with genotype $(x_1,x_2)$. 
	\item[(iv)] $m(x_1,x_2,y_1,y_2,z_1,z_2) \in [0,1]$ : the mating and mutation measure gives the probability that the mating of an individual with genotype $(x_1,x_2)$ with an individual with genotype $(y_1,y_2)$ produces an offspring with genotype $(z_1,z_2)$. It is assumed to satisfy \begin{itemize}
		\item[(a)] for each $\mathbf{x},\mathbf{y} \in \X^2$ $$
		\int\limits_{\X^2} m(\mathbf{x},\mathbf{y},d\mathbf{z}) = 1 \quad \text{ and } \quad \int\limits_{\R^{2N} \setminus \X^2} m(\mathbf{x},\mathbf{y},d\mathbf{z}) = 0.
		$$ Note that since $\lvert \X^2 \rvert < \infty$ this means, that $m(\mathbf{x},\mathbf{y},\cdot)$ is a probability mass function with mass exclusively on $\X^2$.
		\item[(b)] for every $(x_1,x_2),(y_1,y_2),(z_1,z_2) \in \X^2$ the following symmetry properties \begin{align*}
		m(x_1,x_2,y_1,y_2,z_1,z_2) &= m(x_2,x_1,y_1,y_2,z_1,z_2) \\
		m(x_1,x_2,y_1,y_2,z_1,z_2) &= m(x_1,x_2,y_2,y_1,z_1,z_2) \\
		m(x_1,x_2,y_1,y_2,z_1,z_2) &= m(y_1,y_2,x_1,x_2,z_2,z_1) 
		\end{align*}
		The first two properties correspond to the fact, that we do not want to make a difference between the two genotypes of an individual. Both are equally present in the production of the offsprings genotype. Whereas the second property yields that the mating of two individuals has the same probabilities of producing a given pair of genotypes regardless the order of the mating.
	\end{itemize}
\end{itemize}
For simplicity we ignore the existence of sexes and spacial structures within the population. Hence an individual chooses a mate with probabilities only proportional to the birthrate of the partner.
At any point in time $t\geq0$ we consider a finite number $N_t$ of individuals. Denote their genotypes as $(x_1^1,x_2^1), \dotsc, (x_1^{N_t},x_2^{N_t}) \in \X^2$. The population state at time $t\geq0$ is described by the point measure on $\X^2$ $$
\nu_t = \sum\limits_{i=1}^{N_t} \delta_{(x_1^i,x_2^i)}
$$ where $\delta_{(x_1,x_2)}$ is the Dirac measure at $(x_1,x_2) \in \X^2$. 
Let $\langle \nu , f \rangle$ denote the integral of a measurable function $f$ with respect to the measure $\nu$. Then $\langle \nu_t , 1 \rangle = N_t $ and for any $(x_1,x_2) \in \X^2$, the non-negative number $\langle \nu_t, \mathds{1}_{\{(x_1,x_2)\}} \rangle $ is called the density of genotype ${(x_1,x_2)}$ at time $t$. In an abuse of notation we define $$
\langle \nu_t , \mathds{1}_x \rangle \coloneqq \langle \nu_t(x,dy), 1 \rangle + \langle \nu_t(dy,x), 1 \rangle
$$ to be the density of the haplotype $x \in \X$ at time $t$. 
Let $\mathcal{M}\left(\X^2\right)$ denote the set of finite, nonnegative point measures on $\X^2$, equipped with the weak topology, $$
\mathcal{M}\left(\X^2\right) \coloneqq \left\{ \sum\limits_{i=1}^n \delta_{(x_1^i,x_2^i)} \colon n \geq 0 , (x_1^1,x_2^1), \dotsc, (x_1^n,x_2^n) \in \X^2 \right\}
$$ 
An individual with genotype $(x_1,x_2)$ in the population $\nu_t$ reproduces with an individual with genotype $(y_1,y_2)$ at a rate $b(x_1,x_2)\frac{b(y_1,y_2)}{\langle \nu_t , b \rangle}$. The genotype of the offspring is chosen according to the mutation and mating measure $m(x_1,x_2,y_1,y_2,dz_1,dz_2)$.
An individual with genotype $(x_1,x_2)$ in the population $\nu_t$ dies at rate $$
d(x_1,x_2) + \langle \nu_t , c(x_1,x_2,dy_1,dy_2) \rangle
$$
The population process $(\nu_t)_{t \geq 0}$ is defined as a $\mathcal{M}(\X^2)$-valued Markov process with the dynamics described above. These are encoded in the infinitesimal generator $\mathcal{L}$ of the process, which is defined for any bounded measurable function $f:\mathcal{M}(\X^2)\rightarrow \R$ and for all $\nu \in \mathcal{M}(\X)$, by 
\begin{align*}
\left(\mathcal{L}f\right)(\nu) = & \int\limits_{\X^2} b(\mathbf{x}) \int\limits_{\X^2} \frac{b(\mathbf{y})}{\langle \nu , b \rangle} \int\limits_{\X^2} \left(f\left(\nu + \delta_{\mathbf{z}}\right) - f\left(\nu\right)\right) m(\mathbf{x},\mathbf{y},d\mathbf{z}) \nu(d\mathbf{y}) \nu(d\mathbf{x}) \\
& + \int\limits_{\X^2} \left( d(\mathbf{x}) + \int\limits_{\X^2} c(\mathbf{x},\mathbf{y}) \nu(d\mathbf{y}) \right) \left( f\left(\nu - \delta_{\mathbf{x}}\right) - f\left(\nu\right) \right) \nu(d\mathbf{x})
\end{align*}
The first term describes the mating and birth event. The second term describes the death of an individual. We ignore the unnatural fact that an individual can choose itself as a partner to mate as the probability of that event will become negligible as the population size increases.

\begin{remark} Since we assume the model parameters $b,d,c$ take finite, non-negative values, and the trait space $\X^2$ is finite we immediately get the existance and uniqueness of the process. Since if the population is of finite size $n$ and in the state $\nu = \sum\limits_{i=1}^n \delta_{\mathbf{x_i}}$ the total event rate is  $$
	R(\nu) = \sum\limits_{i=1}^n b(\mathbf{x}) + d(\mathbf{x}) + \int\limits_{\X^2} c(\mathbf{x},\mathbf{y}) \nu(d\mathbf{y}) \leq n \left( \max\limits_{\mathbf{x} \in \X^2} \left\{ b(\mathbf{x}) + d(\mathbf{x}) \right\} \right) + n^2 \max\limits_{\mathbf{x},\mathbf{y} \in \X^2} c(x,y) < \infty
	$$ bounded from above as long as the population size is finite. 
\end{remark}

\noindent We see that this is true on finite time intervals as long as we start in a possibly random population with finite mean. 

The trait space is $\mathcal{X} = \{0,1\}^N$ hence every individual is characterized by a $2\times N$ matrix with values in $\{0,1\}$. 
Here zero represents the wild type and a one indicates that (at least one) mutation is present. 
Define the set $\mathcal{D}_N \subset \mathcal{X}^2$ as 
$$
	\mathcal{D}_N \coloneqq \left\{ (x,y) \in \mathcal{X}^2 \colon \exists \, 1 \leq i \leq N \text{ such that } x_i = 1 = y_i \right\} .
$$ 
Then for $x,y,z,w \in \mathcal{X}$ the birth, death and competition rates are given by 
$$
	b(x,y) \coloneqq \bar{b} \mathds{1}_{\mathcal{X}^2\setminus\mathcal{D}_N}(x,y) \quad \text{ and } \quad d(x,y) \coloneqq \bar{d} \quad \text{ and } \quad c(x,y,z,w) \coloneqq \bar{c} 
$$
for some finite $\bar{b},\bar{d}, \bar{c} \in \R_+$. 
Moreover define $\mu>0$ to be the mutation rate per gamete. 
Since usually the number of loci $N$ is big and the mutation rate $\mu$ is small we assume that the number of mutation per birth is Poisson distributed with mean $2\mu$. 
The mutation location then is uniform distributed among all $2N$ possible positions.
During gamete formation, recombination events occur with constant rates. 
Let $r \in [0,1]$ be the probability that a crossover breakpoint occurs between two adjacent gene sequences. 
At these points the genetic information is split and any copy is chosen at uniformly at random to produce the gamete. 
We assume that a crossover breakpoint occurs at any possible cutting point with equal probability $c$, independently of all other points.
Knowing this the probabilities $m(x_1,x_2,y_1,y_2,z_1,z_2)$ can be calculated for any three pairs of genetic information $x_1,x_2,y_1,y_2,z_1,z_2\in\mathcal{X}$ that the paring of $(x_1,x_2)$ with $(y_1,y_2)$ results in $(z_1,z_2)$. 
First define the function $\gamma \colon \{1,\dotsc,N-1\}  \rightarrow \{0,1\} $ that determines weather there is a crossover point between two genes or not in the sense that
\begin{align*}
	\gamma(i) = \begin{cases}
		1 & \text{ if there is a crossover breakpoint between genes } i \text{ and } i+1 \\
		0 & \text{ else.} 
	\end{cases} 
\end{align*}
Then define the choice function $\tau_\gamma \colon \{1,\dotsc,\lVert\gamma\rVert_1+1\} \rightarrow \{1,2\}$ that chooses one of the two copies of each gene segments.
Adding both together we define the function $\phi_{\tau_\gamma}^\gamma \colon \mathcal{X}^2 \rightarrow \mathcal{X}$ that determines the gamete of an individual with crossover points determined by $\gamma \in \{0,1\}^{N-1}$ and chromosome selection $\tau_\gamma \in\{1,2\}^{\lVert \gamma \rVert_1 + 1}$ as 
\begin{align*}
	\phi^\gamma_{\tau_\gamma}(x_1,x_2) =  \left(x^1_{\tau_\gamma(1)},x^2_{\tau_\gamma\left(\gamma(1)+1\right)},\dotsc, x^N_{\tau_\gamma\left(\gamma(1)+\dotsc+\gamma(N-1) + 1 \right)} \right)_{k=1,\dotsc,N}
\end{align*}
Then we can define the mating and mutation probabilities in a general setting as 
\begin{align*}
	m(\mathbf{x},\mathbf{y},d\mathbf{z}) 
	& = \sum\limits_{\gamma_x,\gamma_y \in\{0,1\}^{N-1}} r^{\lVert \gamma_x \rVert_1 + \lVert \gamma_y \rVert_1} (1-r)^{2N-2-\lVert \gamma_x \rVert_1 - \lVert \gamma_y \rVert_1} \frac{1}{2^{\lVert \gamma_x \rVert_1 + \lVert \gamma_y \rVert_1+2}} \\
	&\phantom{=} \times \sum\limits_{\substack{\tau_x \in \{1,2\}^{\lVert \gamma_x\rVert_1 + 1} \\ \tau_y \in \{1,2\}^{\lVert \gamma_y\rVert_1 + 1}}} \sum\limits_{k=0}^\infty \frac{(2\mu)^k}{k!}e^{-2\mu} \frac{1}{Z_k} \sum\limits_{m\in\Diamond^{2N}_k} \delta_{\left(\left(\phi_{\tau_x}^{\gamma_x}(\mathbf{x}),\phi_{\tau_y}^{\gamma_y}(\mathbf{y})\right) + m \right) \wedge 1} (d\mathbf{z})
\end{align*}
where $\Diamond_k^{2N} \coloneqq \left\{ m \in \mathbb{N}_+^{2N} \colon m_1 + \dotsc +m_{2N} = k \right\}$ is the set of all lattice vectors in $\mathbb{N}_+^{2N}$ with one norm equal to $k$, moreover $ Z_k = \sum_{j=0}^{2N} \binom{2N}{j} p_j(k) $ is the size of the set $ \Diamond_k^{2N} $ and where $p_j(k)$ is the number of partitions of $k$ into exactly $j$ parts.
For notational reasons define for $x\in\R^{2N}$ and $k\in\R$ the component wise maximum as $x \wedge k \coloneqq (x_1\wedge k,\dotsc,x_{2N}\wedge k)$.

\subsubsection{Results} 
In the initial equilibrium where $c_0$ is present, certain parameter combinations of $N$ and $\mu$ may result in a total population size lower than the carrying capacity.
However, all relative statistics such as prevalence and mutation burden remain comparable to those of the constant size model.
The diminished population sizes associated with certain parameter combinations engender higher relative fluctuations, thereby favouring the extinction of the least loaded class due to natural fluctuations.
The extinction of $c_0$ in this scenario triggers a rapid escalation in mutation burden and prevalence.
Unlike the constant size model, the remaining healthy individuals cannot effectively manage the swiftly rising prevalence, leading to the population's rapid extinction.
By modifying the model to maintain a constant overall birthrate evenly distributed among all healthy individuals, we can replicate the exact dynamics observed in the Wright-Fisher model.

\subsection{Remark on Recombination}\label{sec:rem_rec}
In this work, recombination is implemented as a two-step process.
First, potential crossover breakpoints are determined on the genome, and then the respective gene segments are selected with equal probability between the maternal and paternal genome.
As a result, on average, there are $\frac{Nr}{2}$ \emph{true} crossover breakpoints.
Since a \emph{true} crossover breakpoint occurs only when different origins are chosen for two consecutive genes, which happens at each potential breakpoint with probability $\frac{1}{2}$.
An alternative implementation, often found in the literature, combines these two processes into one.
There, crossover breakpoints are also determined for each individual with equal probability between each gene, but these automatically lead to switching between the maternal and paternal genome, or vice versa.
Hence, only one starting genome is selected (maternal or paternal), and then switching occurs automatically at each breakpoint.
Both implementations are equivalent.
However, since the selection process is already included in the latter, the recombination rate - which is the probability of occurrence of a crossover breakpoint between neighbouring genes - ranges from 0 to $\frac 1 2$ instead of $r \in [0,1]$ as in the former.
Accordingly, the case of full recombination, which we consider in this work, is equivalent to a recombination rate of $r= \frac 1 2$ for models with the second implementation of recombination.

\bibliographystyle{abbrv}
\bibliography{Referenzen}

\end{document}